\documentclass[%
 reprint,
nofootinbib,
 amsmath,amssymb,
 aps,
prb,
floatfix,
]{revtex4-1}
\usepackage{hyperref}
\usepackage{pgfplots}
\usepackage{graphicx}% Include figure files
\usepackage{dcolumn}% Align table columns on decimal point
\usepackage{bm}% bold math
\usepackage{wrapfig}
\usepackage{graphicx}
\usepackage{verbatim}
\usepackage{float}
\begin{document}

%\preprint{APS/123-QED}

\title{Origins of magnetic field-dependent open-circuit voltage hysteresis driven by transverse charge current in ferromagnet/normal metal structures}% Force line breaks with \\

\author{Christos Tengeris}
 %\altaffiliation[Also at ]{Department of Physics, Center for Nanophysics and Advanced Materials, University of Maryland, College Park, Maryland 20742, USA}%Lines break automatically or can be forced with \\
 %\email{Second.Author@institution.ed}
\affiliation{
 Department of Physics, University of Maryland, College Park, Maryland 20742, USA}%\textbackslash\textbackslash


%\affiliation{
%\affiliation{
 %Third institution, the second for Charlie Author
%\affiliation{%
 %This line break forced with} %\textbackslash\textbackslash

\date{February 24, 2020}
           
\begin{abstract}
Recent experimental work on Au thin films demonstrated signs of charge current-induced spin polarization through open circuit voltage measurements. In this study, we are investigating the underlying mechanism(s) that induces this measured signal in the Au devices. We determine the theoretically expected spin polarization from both Rashba-Edelstein effect and bulk spin Hall effect. The discrepancy in the scaling of the measured signal as a function of the thickness of the Au thin film in the two cases is our key to differentiate between the two effects when compared to experimental data. Experiments show reversal of spin polarization at a critical thickness which reveals the presence of multiple spin polarization mechanisms. Characteristics of both Rashba-Edelstein and spin Hall effects are observed in different thickness regimes. In addition, we study the magnetoresistance of the same Au samples, which reveal the presence of weak anti-localization (WAL) at low temperatures for the low-thickness samples. More interestingly, it is revealed that the open circuit voltage difference and magnetoresistance due to WAL have very similar scaling with film thickness and temperature, suggesting the crucial importance of spin-orbit interaction in understanding the phenomenon.

\end{abstract}

\pacs{Valid PACS appear here}% PACS, the Physics and Astronomy
\maketitle

%\tableofcontents

\section{\label{sec:level1}Introduction}
In the recent past, there have been several experiments to demonstrate signs that spin polarization can be induced by driving a charge current through a topological insulator \cite{li2016direct,ando2014electrical,dankert2015room,tang2014electrical,tian2015electrical,lee2015mapping}. In these experiments, an electric current is driven in the plane of a thin film of topological insulator, and the voltage on a ferromagnetic metal contact is used to deduce the presence of a nonzero in-plane spin polarization, oriented perpendicular to the driven current. This observation of current-induced spin polarization was attributed to presence of electronic surface states that bridge the energy gap of these otherwise insulating materials and form `helical' bands. In these helical bands, spin states and momentum states are uniquely related (spin-momentum locking)\cite{hsieh2009tunable}. As a consequence, an ensemble spin polarization is induced when an electric field creates an ensemble electron momentum imbalance. These observations were followed by a similar experiment where the same effect was observed in devices with topologically trivial Au instead of a topological insulator\cite{Li_PRB16}. The present document is the record of an investigates into the origins of this current-induced spin polarization in Au.%

This paper is organized as follows. In the second section, we present the main two theoretical models used to explain the presence of current induced spin polarization in thin films of Au. These are the Rashba-Edelstein effect \cite{bychkov1984properties} and the bulk spin Hall effect. Both mechanisms are a consequence of the strong spin orbit coupling (SOC) in Au. In the third section we present our measurements on CoFe/Au devices that show how the spin polarization signal depends on temperature and the thickness of thin Au films. This data reveal a surprising reversal of spin polarization direction at a critical thickness of the Au film, which suggests the presence of multiple competing mechanism that generate spin polarization. Magnetoresistance, revealing weak anti-localization, is also investigated in the same Au/CoFe devices. In the final section, we list some conclusions from this study.

\section{Theory}
\begin{figure}
%\vspace{-10pt}	
%\centering
    \includegraphics[width=0.4\textwidth]{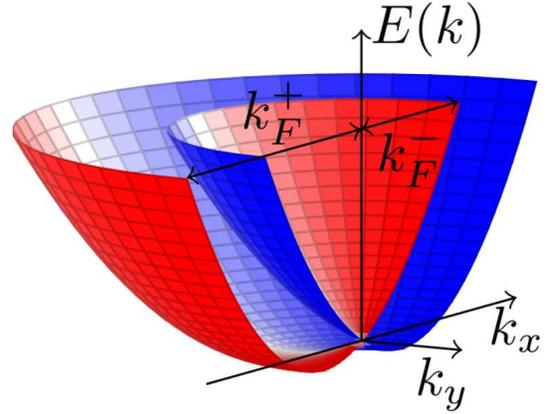}%{Rashba.png}
	\caption{Dispersion relation in a system with Rashba spin orbit interaction. Color grading represents spin states with red being spin up and blue being spin down. The spin orientation is consistent with spin-momentum locking where spins are oriented in plane and perpendicular to $\vec{k}$.
	\label{fig:E_f(k)}}
    \vspace{-10pt}
\end{figure}
The two possible mechanisms we propose to explain the presence of this open-circuit  voltage hysteresis signal that is observed in Au/CoFe devices and is presumably due to spin polarization in Au are the Rashba-Edelstein effect at the substrate-Au interface and/or the bulk spin Hall effect (SHE). In this section, we use theoretical models to calculate the resulting spin polarization after applying a constant charge current to a thin layer of Au for both scenarios. 

\subsection{Current-induced spin polarization due to Rashba dispersion}
The Rashba-Edelstein effect is a consequence of the joint effects of inversion symmetry breaking and spin orbit interaction (SOI) in a system. In our case, the inversion symmetry breaking comes from the mismatch of work function in Au and the electron affinity in SiO$_2$ (substrate). This creates a very sharp electrical potential difference at the interface where the two materials meet, and associated out-of-plane electric field. In materials like Au where the SOC is strong, such electric fields give rise to the Rashba-Edelstein effect. Therefore, an extra Hamiltonian contribution due to this effect should be considered when dealing with such systems \cite{rashba1959symmetry,rashba2015symmetry,bychkov1984properties}
\begin{equation}
	H_{Rashba}= \alpha (\overrightarrow{\sigma} \times \overrightarrow{p})\cdot \hat{z}
\label{eq:H_rash}      
\end{equation}
where $\alpha = \frac{g\mu _b E_o}{2mc^2}$ and $E_o$ is the magnitude of the out-of-plane electric field. The above Hamiltonian term is responsible for the splitting of the two otherwise degenerate spin sub-bands, such as the case for parabolic bands shown in Fig. \ref{fig:E_f(k)}. We can see that after the spin splitting, we obtain two separate helical bands where the spin of the electronic states is related to their quasi-momentum $\vec k$. %* momentum or wave vector

Because of the structure of this spin-split conduction band, it is possible in principle to generate spin polarization just by applying an electric field to such a system. The applied field will lead to the asymmetric occupation of these bands in a way that positive momentum states are going to be populated more than negative momentum states along the direction of the applied electric field. Since momentum state and spin states are correlated, a non-equilibrium spin polarization (oriented perpendicular to the charge current) is produced. This charge current induced spin polarization can be calculated as follows.   
\subsubsection{Dispersion}

The energy of the electron states is given by
\begin{equation}
	\centering
	E(k_x,k_y) = \frac{\hbar ^2}{2m}(k_x^2+k_y^2)\pm \lambda \hbar \sqrt{k_x^2+k_y^2},
	\label{eq:E}
\end{equation}
where $\lambda$ is a constant proportional to the spin splitting due to the Rashba effect\cite{bychkov1984properties}. A surface plot of this expression can be seen in Fig. \ref{fig:E_f(k)}. The positive sign in \eqref{eq:E} corresponds to the surface above the point $\vec{k}=(0,0)$ and the minus sign corresponds to the surface below that point. Note that spins have opposite helicities on these two surfaces.  
%We consider here only the case that $E_F>0$. 
The wavevectors at the two Fermi surfaces $k_F^{\pm}$ can be obtained by solving the quadratic equation $\frac{\hbar ^2 k^2}{2m} + \hbar \lambda k =E_F$ to get 
\begin{equation}
\centering
k_F^\pm=\sqrt{\frac{2mE_F}{\hbar ^2}+(\frac{m\lambda}{\hbar})^2}\pm \frac{m \lambda}{\hbar}=k_0\pm \frac{m \lambda}{\hbar}.\label{eq:kFpm}  
\end{equation}

\subsubsection{Total current}
The first step to calculate the current-induced spin density polarization is to evaluate the total (charge) current that is flowing in such a material when an electric field is applied along the $x$ direction. The expression for the current is given by
\begin{equation}
	\centering
	j = -e\int v_x g(\vec{k}) \frac{d^2k}{(2\pi )^2},
	\label{eq:j_1}
\end{equation}
where $g(\vec{k})$ is the anti-symmetric part of the (distorted by $\mathcal{E}_x$) occupation function, given by $g(\vec{k})= \Delta k_x \frac{df}{dk_x}$ where $\Delta k_x = \frac{e\tau \mathcal{E}_x}{\hbar}$ and $\frac{df}{dk_x}= -\delta (k-k_F)\cos\theta$. The (group) velocity
\begin{align}
	\centering
	v _x^\pm &= \frac{d}{\hbar dk_x} E(k_x,k_y)=\frac{1}{\hbar}\left[\frac{\hbar ^2}{2m}(2k_x)\pm \lambda \hbar \frac{k_x}{\sqrt{k_x^2+k_y^2}}\right] \nonumber \\ 
    &= \frac{\hbar k_x}{m} \pm \frac{\lambda k_x}{k} = (\frac{\hbar k}{m} \pm \lambda)\cos\theta
	\label{eq:u_x}
\end{align}
where $k_x=k\cos\theta$. Now substitute into Eq.\eqref{eq:j_1}
to get

\begin{align}
	\centering
	j& =\frac{e}{(2 \pi)^2} \int \left[(\frac{\hbar k}{m}-\lambda)\delta (k-k_F^+)+(\frac{\hbar k}{m}+\lambda)\delta (k-k_F^-)\right] \nonumber \\ 
    &\cos^2\theta\Delta k_x kdkd\theta \nonumber \\ 
    & = \frac{e}{4\pi} \left[(\frac{\hbar}{m}k_F^+-\lambda)k_F^++ (\frac{\hbar}{m}k_F^-+\lambda)k_F^-\right] \Delta k_x \nonumber \\ 
    &=\frac{e\hbar k_0^2}{2\pi m}\Delta k_x.
	\label{eq:j_tot}
\end{align}
Note that when $\lambda=0$ such that $k_0=k_F=\sqrt{2mE_F}/\hbar$ and the density $n=\frac{k_F^2}{2\pi}$, this gives $j =\frac{ne^2\tau}{m}\mathcal{E}_x$ (the standard Drude result).

\subsubsection{Spin density and polarization}
The next step in this calculation is to evaluate the spin density of the occupied states. Spin density is given by
\begin{align}
	\centering
	n_{\uparrow} & = \int_0^{2\pi} \int_0^{\infty} \left[(f_0 (\vec{k})+\Delta k_x \delta (k-k_F^+))\cos^2\frac{\theta}{2} \right.\nonumber \\
    &+\left. (f_0 (\vec{k})+\Delta k_x \delta (k-k_F^-)\sin^2\frac{\theta}{2} \right] kdkd\theta \nonumber \\
n_{\downarrow} & = \int_0^{2\pi} \int_0^{\infty}\left[ (f_0 (\vec{k})+\Delta k_x \delta (k-k_F^+))\sin^2\frac{\theta}{2} \right. \nonumber \\ 
&+ \left. (f_0 (\vec{k})+\Delta k_x \delta (k-k_F^-)\cos^2\frac{\theta}{2}\right] kdkd\theta \nonumber
\end{align}
yielding
\begin{align}
n_{\uparrow/\downarrow} &= \frac{1}{8 \pi}[({k_F^+}^2\pm k_F^+ \Delta k_x)+ ({k_F^-}^2 \mp k_F^-\Delta k_x)] \nonumber \\ 
&= \frac{1}{8 \pi}\left[({k_F^+}^2 + {k_F^-}^2) \pm (k_F^+ - k_F^-) \Delta k_x)\right].
	\label{eq:n}
\end{align}
From the definition of the current-induced spin density polarization we get:
\begin{equation}
	\centering
	P_{CI} = \frac{n_\uparrow - n_ \downarrow}{n_\uparrow + n_ \downarrow} = \frac{k_F^+ - k_F^-}{{k_F^+}^2 + {k_F^-}^2} \Delta k_x
	\label{eq:Pci_1}
\end{equation}
Note that when $k_F^-=0$ this reduces to the result for Dirac cone  \cite{Li_PRB16}. After substitution of $\Delta k_x$ obtained from Eq. \ref{eq:j_tot}, we have 
\begin{equation}
	\centering
	P_{CI} =  \frac{2\pi m}{e\hbar } \frac{k_F^+ - k_F^-}{k_0^2({k_F^+}^2 + {k_F^-}^2)}j.
	\label{eq:Pci_2}
\end{equation}
Using the definition of $k_F^{\pm}$ from Eq. \eqref{eq:kFpm}, we can express this quantity as
\begin{equation}
	\centering
	P_{CI} =2\pi\left[ \frac{ \lambda m^2 }{e \hbar^2k_0^2 \left(k_0^2 + (\frac{m \lambda }{\hbar })^2\right)}\right]j.
	\label{eq:Pci_3}
\end{equation}

\subsubsection{Comparison to approximation}
Eqn. 2 of Ref. \onlinecite{Li_PRB16} suggests $P_{CI}^\pm=\frac{\Delta k_x}{k_F^\pm}$ for each [inner(-) or outer(+)] Fermi surface of the Rashba dispersion. We must average their contributions, weighted by density:
\begin{align}
P_{CI}=\frac{P_{CI}^+n^+-P_{CI}^-n^-}{n^++n^-},
\label{eq:Pdirac}
\end{align}
Using $n^\pm={k_F^\pm}^2/4\pi$, we then recover Eq. \ref{eq:Pci_1} exactly.

Of course the above result gives the spin polarization at the bottom interface where the Au layer and the SiO$_2$ substrate meet. The spin polarization at the top interface between Au and CoFe\textemdash where the spin detection takes place\textemdash is generally different. Assuming a simple diffusion model, the intensity of the detected spin polarization is going to be $P=P_{CI}e^{-L/\lambda_{\sigma} }$, where $L$ is the total thickness of the thin Au film and $\lambda_{\sigma} $ is the spin diffusion length in Au.

\subsection {Current-induced spin polarization due to bulk Spin Hall Effect}
Another phenomenon that can explain the generation of spin polarization from sourcing charge current is the bulk spin Hall effect (SHE). In SHE, spin-dependent scattering of the carrier makes electrons of opposite spin to move in opposite directions perpendicular to the applied current and therefore accumulate at opposite surfaces of the conducting channel. We can calculate the spin polarization predicted from this mechanism by solving the coupled drift-diffusion equation with finite spin lifetime\cite{yu2002electric}.

\iffalse
\subsubsection{No spin relaxation}
Spin-up and spin-down are completely decoupled, and the steady-state solution to 
$\frac{dn}{dt}=D\frac{d^2n}{dx^2}-v\frac{dn}{dx}$
is $n=A\exp(xv/D)+B$. Insulating boundary conditions ($D\frac{dn}{dx}|_{x=-L/2}-vn(-L/2)=0$ and $vn(L/2)-D\frac{dn}{dx}|_{x=L/2}=0$) give $B=0$. Normalization via 
$\int_{-L/2}^{L/2}n(x)dx=N$ gives $A=\frac{v}{D}\frac{N}{2\sinh(vL/2D)}$. Therefore
\begin{align}
n_\uparrow(x)=\frac{v}{D}\frac{N}{2\sinh(\frac{vL}{2D})}\exp(\frac{v}{D}x)
\end{align}

For opposite spins the sign of velocity $v$ is reversed, yielding
\begin{align}
n_\downarrow(x)=\frac{v}{D}\frac{N}{2\sinh(\frac{vL}{2D})}\exp(-\frac{v}{D}x)
\end{align}

The polarization is given by
\begin{align}
P(x)=\tanh(\frac{vx}{D})
\end{align}
which gives $P(-L/2)=-\tanh(\frac{vL}{2d})$ and $P(L/2)=\tanh(\frac{vL}{2d})$ at the boundaries. Note that this function behaves as $Lv/2D$ for small values relevant to experiments. No nonlinearity exists as $v\rightarrow 0$.
\fi

\begin{align}
D\frac{d^2n_\uparrow}{dx^2}-v\frac{dn_\uparrow}{dx}-\frac{n_\uparrow-n_\downarrow}{\tau}&=0,\\
D\frac{d^2n_\downarrow}{dx^2}+v\frac{dn_\downarrow}{dx}-\frac{n_\downarrow-n_\uparrow}{\tau}&=0.
\end{align}
Where $n_{\uparrow,\downarrow}$ are the carrier densities for spin up and down electrons, $D$ is the diffusivity and $u$ is the Hall velocity and is proportional to the drift velocity and Hall angle.
Adding and subtracting, we obtain differential equations for the total $n=n_\uparrow+n_\downarrow$ and difference $\Delta n=n_\uparrow-n_\downarrow$,
\begin{align}
D\frac{d^2\Delta n}{dx^2}-v\frac{dn}{dx}-\frac{2\Delta n}{\tau}&=0;\\
D\frac{d^2n}{dx^2}-v\frac{d\Delta n}{dx}&=0.
\end{align}
It is apparent that solutions to this system can be written $n=A\cosh(a x)+C$ and $\Delta n = B\sinh(a x)$; substitution then yields the algebraic relationship $B=ADa/v$ and $a=\pm\sqrt{\frac{v^2}{D^2}+\frac{2}{D\tau}}$. The latter can be expanded in the limit of small $v$ as 
\begin{align}
|a|\approx \sqrt{\frac{2}{D\tau}}(1+\frac{v^2\tau}{4D}) \quad (v\rightarrow 0).
\end{align}
Conversely, in the limit of long lifetime $\tau$, it can be expanded as  
\begin{align}
|a|\approx \frac{v}{D}(1+\frac{D}{v^2\tau}) \quad (\tau\rightarrow \infty).
\end{align}
Insulating boundary conditions at $x=\pm L/2$ ($D\frac{dn}{dx}|_{x=-L/2}-vn(-L/2)=0$ and $vn(L/2)-D\frac{dn}{dx}|_{x=L/2}=0$) can be combined to dictate at $x=L/2$

\begin{align}
vn-D\frac{d\Delta n}{dx}=0,%,\\v\Delta n-D\frac{dn}{dx}=0.
\end{align}
%Both give the relationship between $C$ and $A$ as
which gives the relationship between $C$ and $A$ as
\begin{align}
C=A\cosh(aL/2)\left[ \left(\frac{Da}{v}\right)^2-1\right],
\end{align}
which clearly vanishes as $\tau\rightarrow\infty$ because $|a|\rightarrow \frac{v}{D}$. The spin polarization, 
\begin{align}
P(x)&=\frac{\Delta n(x)}{n(x)}=\frac{D}{v}\frac{a\sinh(ax)}{\cosh(ax)+C/A} \nonumber \\
&= \frac{D}{v}\frac{a\sinh(ax)}{\cosh(ax)+\cosh(aL/2)((\frac{Da}{v})^2-1)}
\label{eq:Pofx}
\end{align}
asymptotically approaches $\tanh(vx/D)$ in this relaxation-free limit. 
%We can verify this solution with numerical finite-differences simulation; see figure \ref{eq:SHE_pol_lifetime}. 

\begin{figure}
\centering
%\begin{subfigure}
%\centering
\includegraphics[width=0.4\textwidth]{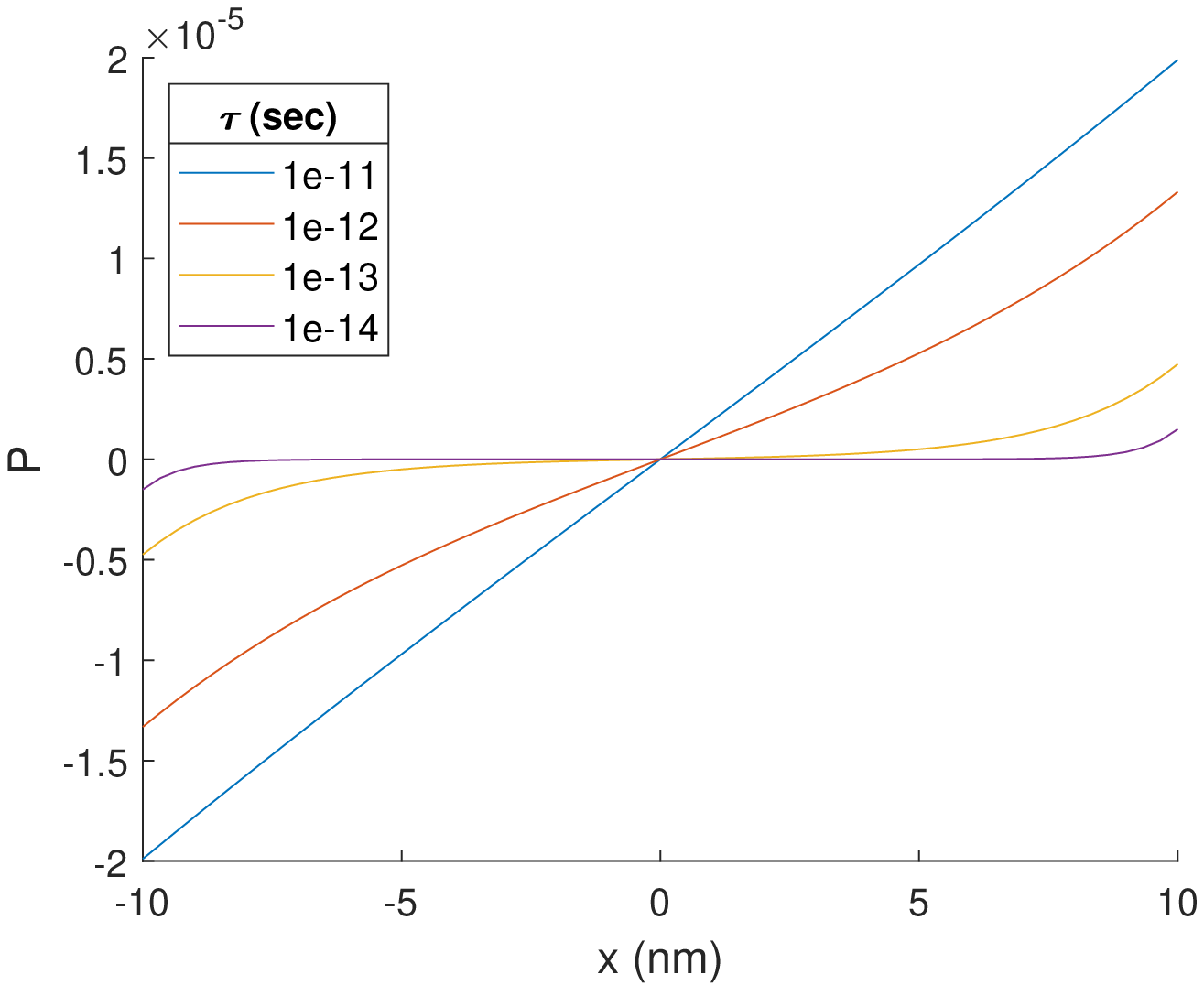} 
%\caption{} \label{fig:sub:P_x}
%\end{subfigure}
%\begin{subfigure}
%\centeringc
\includegraphics[width=0.4\textwidth]{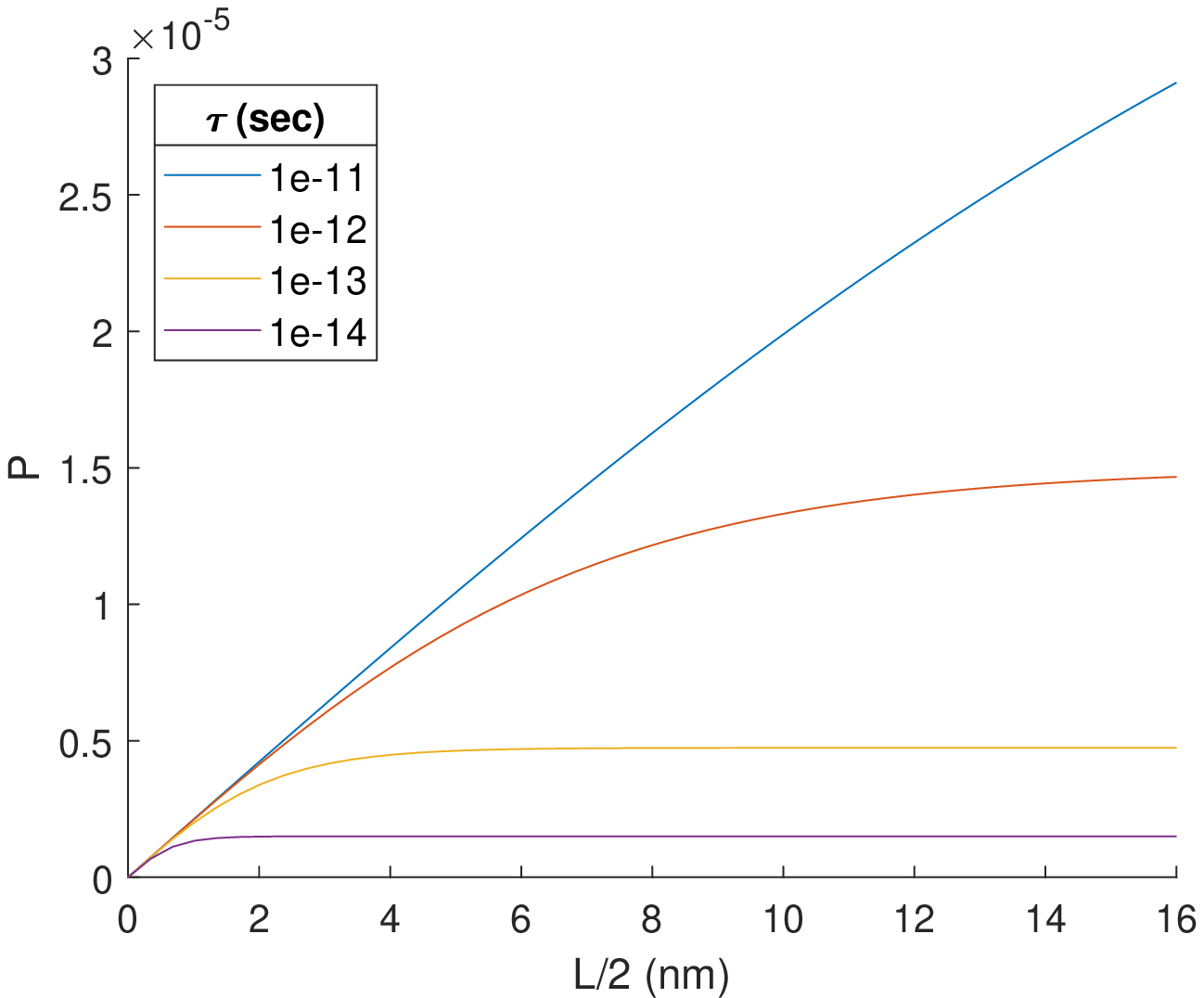} 
%\caption{} \label{fig:sub:P_L2}
%\end{subfigure}
 \caption{Numerical calculations of spin polarization using Eqs. \ref{eq:Pofx} and \ref{eq:PL2} for various values of $\tau$. Upper panel shows the magnitude of spin polarization inside the Au layer at various points along the out of plane direction starting from the Au/SiO$_2$ interface to the Au/CoFe interface. The multiple lines correspond to different values of spin relaxation time in Au. Lower panel shows the magnitude of spin polarization at the Au/CoFe (the point of the measurement) as a function of spin relaxation time. The parameter used for the numerical calculations are: $D=10^{-4}$ m$^2$/s, $u=2.12*10^{-1}$ m/s and $L= 20$~nm.}
 \label{pl:Polarizaion}
\end{figure}

Note that we can write the boundary spin polarization as
\begin{align}
P(\frac{L}{2})=\frac{\Delta n(\frac{L}{2})}{n(\frac{L}{2})}=\frac{v}{Da}\tanh(\frac{aL}{2}).
\label{eq:PL2}
\end{align}
Figure \ref{pl:Polarizaion} shows plots of Eqs. \ref{eq:Pofx} and \ref{eq:PL2} for different values of the spin relaxation constant. 

To lowest order in $1/\tau$, we have
\begin{align}
P(x)\approx \frac{(1+\frac{D}{v^2\tau})\sinh(ax)}{\cosh(ax)+\frac{2D}{v^2\tau}\cosh (vL/2D)},
\label{P_x}
\end{align}
and the polarization at the boundary is $|P(L/2)|\approx(1-\frac{D}{v^2\tau})\tanh(vL/2D)$. In realistic systems, $vL/D\ll 1$ so we have in linear response $\approx \frac{vL}{2D}-\frac{L}{2v\tau}$.
In the limit of short lifetime $\tau\ll 2D/v^2$, $a\approx\sqrt{\frac{2}{D\tau}}$. Then, 
\begin{align}
P(L/2)\approx\sqrt{\frac{\tau v^2}{2D}}\tanh(\frac{L}{\sqrt{2D\tau}}).\qquad(\tau\rightarrow 0)
\label{P_L2}
\end{align}
Variations as a function of thickness $L$ are then expected to be nonlinear only in the diffusion-length regime. 

%\subsection{\label{sec:level2}Second-level heading: Formatting}
%\subsubsection{Wide text (A level-3 head)}
%\subsection{\label{sec:citeref}Citations and References}
%\subsubsection{Citations}

\section{Experimental results}%Thickness dependence of spin polarization and experimental results and discussion
\begin{figure}
	\centering
    \includegraphics[scale=0.12]{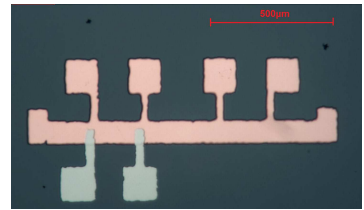}
    \caption{A top down view of a typical Au/CoFe device
    \label{device_pic}}
\end{figure}

\begin{figure}
\includegraphics[width=0.45\textwidth]{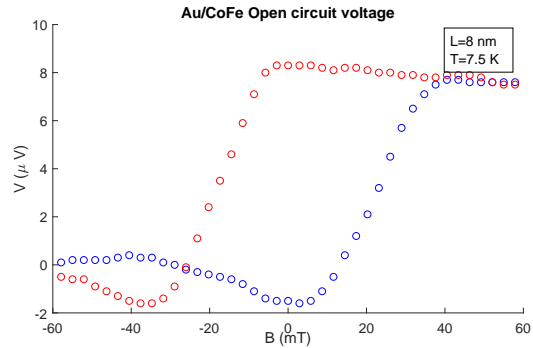}
    \caption{The open circuit voltage (raw data) between Au and CoFe as a function of the in plane magnetic field. This voltage follows a hysteresis curve. The amplitude of the hysteresis loop $\Delta V$ is directly proportional to the spin polarization at Au/CoFe interface that points along the direction of the external magnetic field.}
    \label{fig:V_B}
\end{figure}

As it can be seen from the theoretical analysis of the two candidate mechanisms, the scaling of the observed signal with the thickness of the Au thin film is different in each case. Therefore it is expected that experiments on devices of various thicknesses of Au thin films will be able to help us differentiate between the two mechanisms.
We have prepared samples with different thickness of Au layer from 8nm to 20nm. The devices consist of an Au layer deposited by thermal evaporation followed by the deposition of CoFe magnetic contacts using e-beam evaporation. A top view of the complete device is shown in Figure \ref{device_pic}.
The measurement process follows closely the one in Ref. \onlinecite{Li_PRB16}. In short, a charge current was sourced \footnote{The charge current density was kept constant for all measurements at $j=2\times 10^9$ A/m$^2$. For the sample geometry in this experiment, this corresponds to a current of $200 \mu$A per nm of thickness in the Au thin film.} through the thin Au film, and simultaneously the open circuit voltage between the Au and CoFe is measured as an in plane (and perpendicular to the current) magnetic field is swept in the range of $\pm 60$ mT to control FM magnetization. A typical data set from such measurements is plotted in Figure \ref{fig:V_B}.

\begin{figure}
% \centering
% \begin{subfigure}[H]{0.5\textwidth}
% \centering
\includegraphics[width=0.5\textwidth]{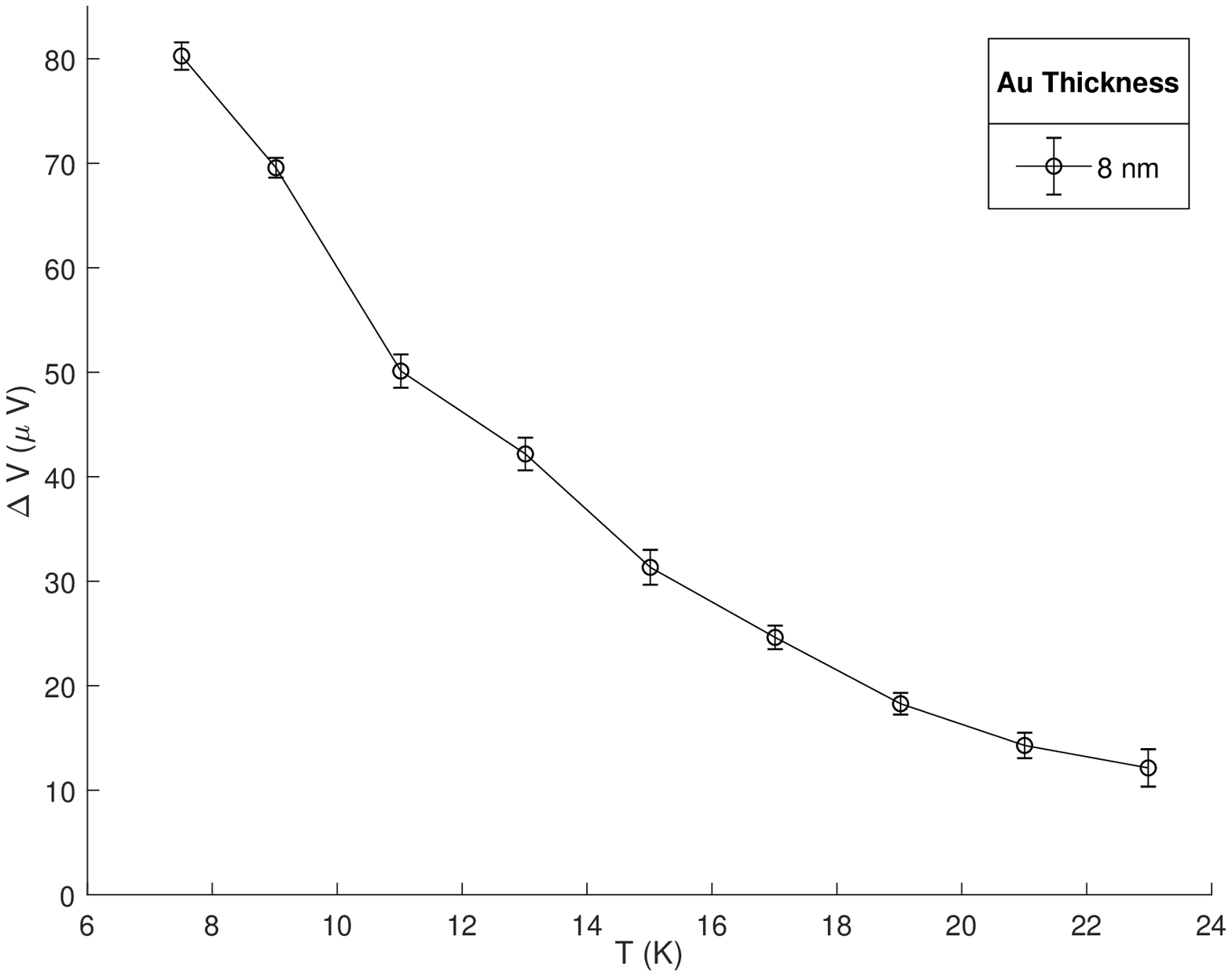} 
% \caption{} \label{fig:l-8}
% \end{subfigure}

% \begin{subfigure}[h]{0.5\textwidth}
% \centering
\includegraphics[width=0.5\textwidth]{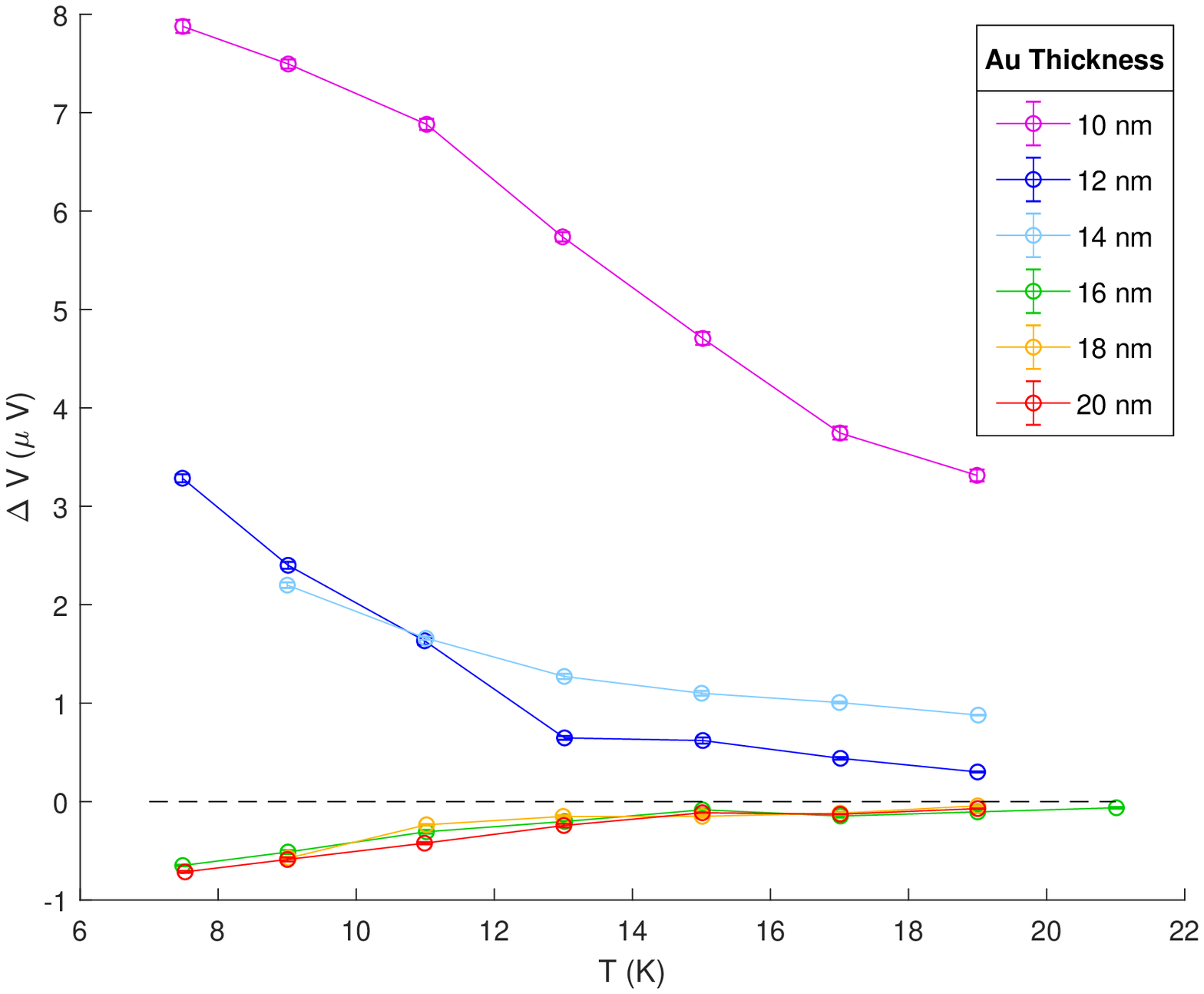} 
% \caption{} \label{fig:l-10-22}
% \end{subfigure}

 \caption{Spin signal $\Delta V$ as a function of temperature. Data are plotted in two different plots due to the much larger signal from the thinnest device.
 \label{pl:DV_L}}
\end{figure}

We can extract the amplitude of the open-circuit voltage hysteresis ($\Delta V$)---defined as the difference in the voltage between Au and CoFe at the two extreme values of the magnetic field---for each one of the samples for various temperatures. The results are presented in Figure \ref{pl:DV_L}. It can be seen that $\Delta V$ decreases monotonically with increased temperature. More interesting though is the scaling of $\Delta V$ with the thickness of the thin film of Au. The plots show a drop in the value of $\Delta V$ as thickness increases up to some critical thickness (14nm-16nm) where we observe reversal of spin polarization for thicker devices. It is also observed that for thicknesses 16nm, 18nm and 20nm, the magnitude of $\Delta V$ is almost identical. 

\begin{figure}
\includegraphics[width=0.5\textwidth]{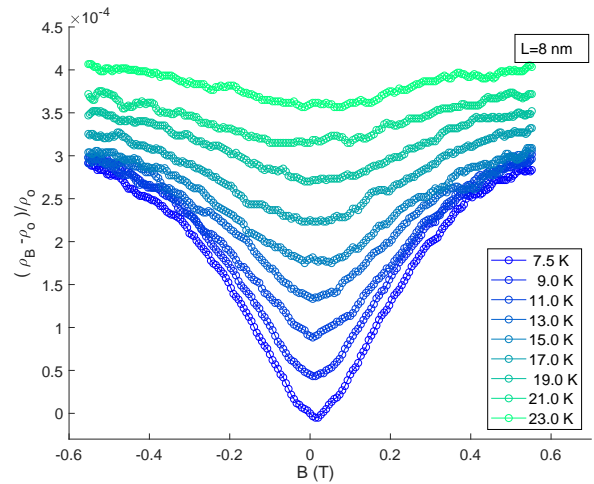}
\caption{Relative deviation from zero field resistivity as a function of out of plane magnetic field. Data reveal the presence of weak anti-localization at low temperatures.}
\label{fig:MR}
\end{figure}

\begin{figure*}
% \centering
% \begin{subfigure}
\includegraphics[scale=0.5]{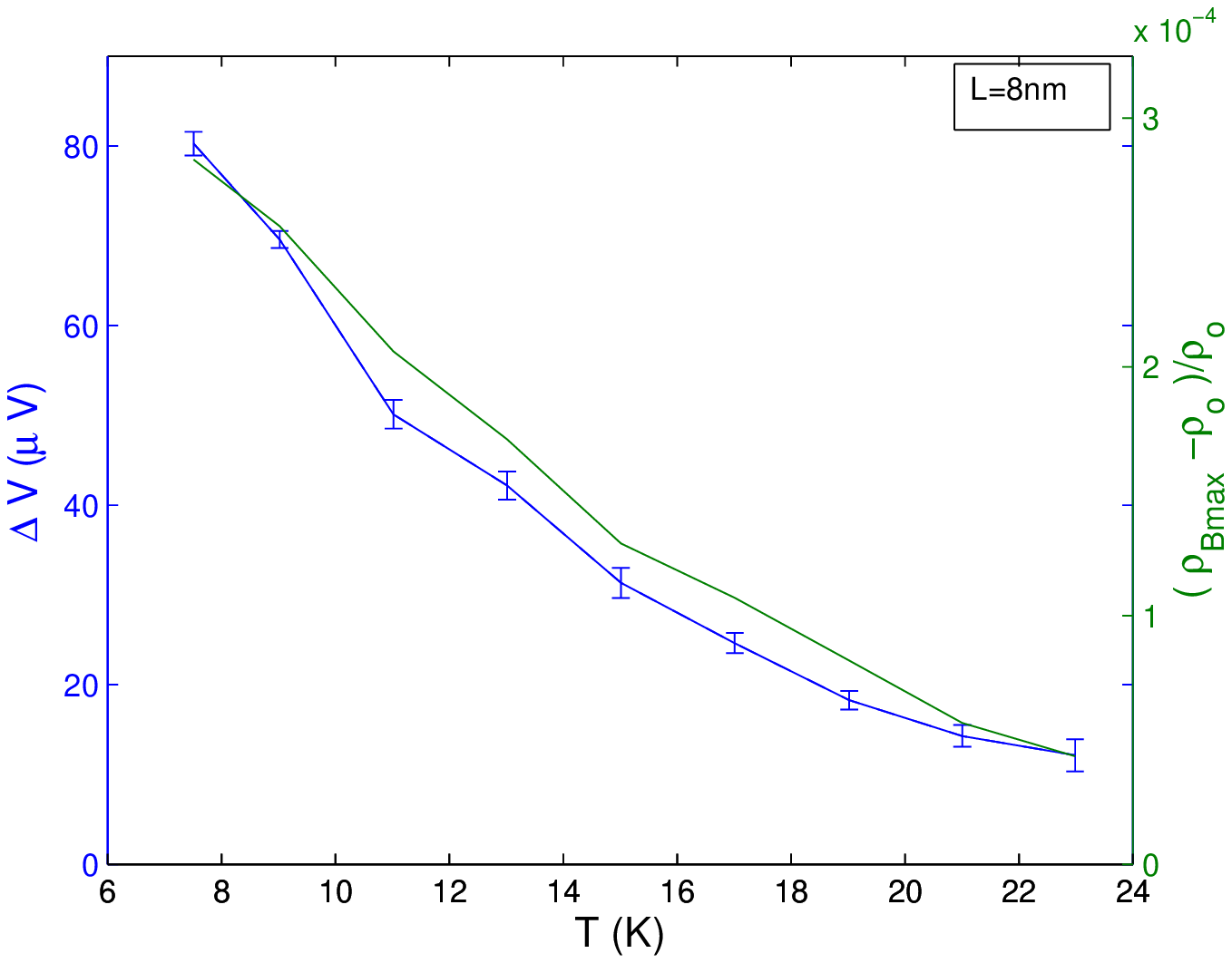} 
\includegraphics[scale=0.5]{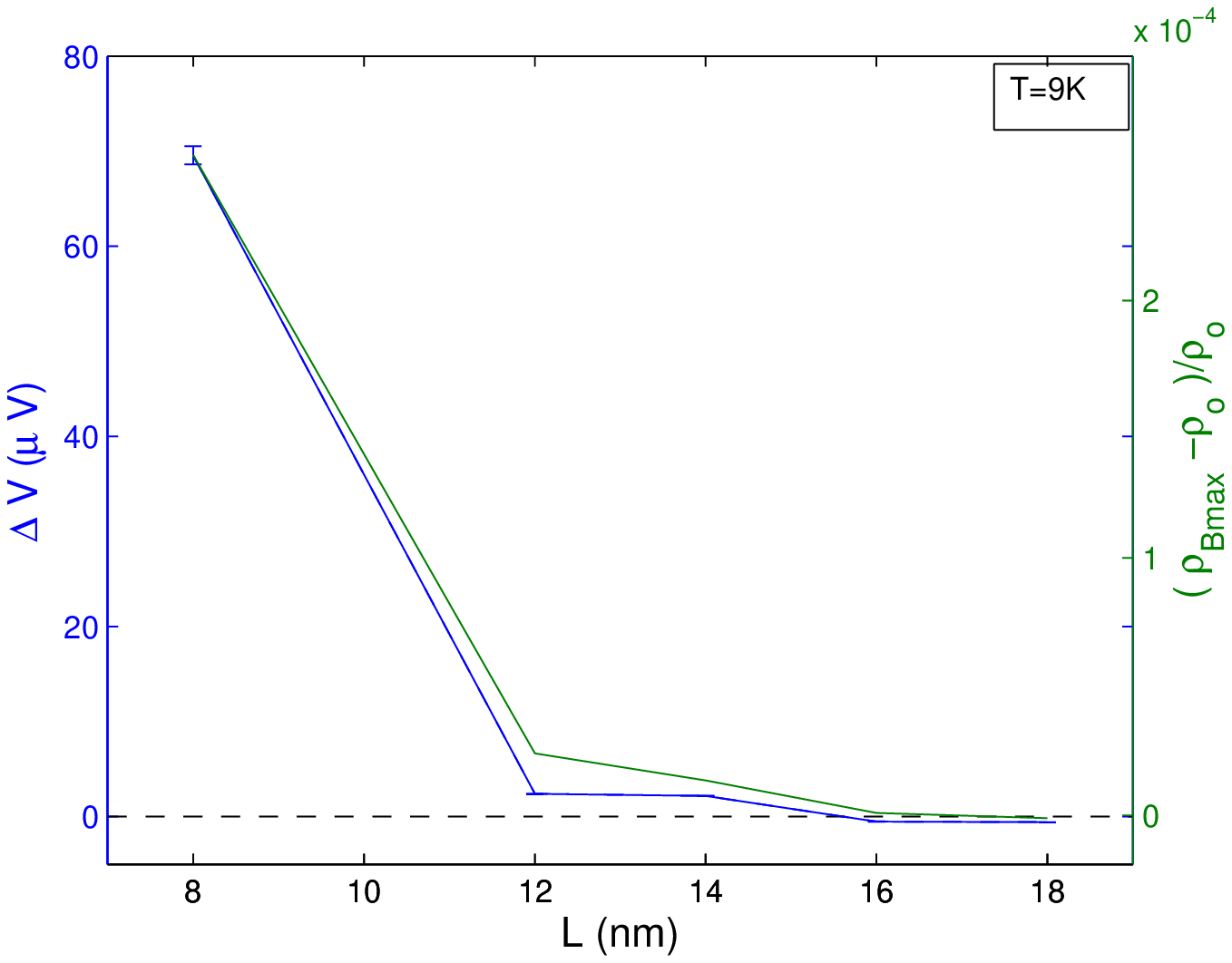} 
%\caption{ \label{fig:sub:MR_L}}
%\end{subfigure}
\caption{Spin signal $\Delta V$ and the magnetoresistance are plotted in the same graph both as a function of temperature (left panel) and device thickness (right panel) in an attempt to investigate the similarity in scaling with these parameters.
\label{fig:DV_MR}}
\end{figure*}

We also perform magnetoresistance (MR) measurements on the same samples. For each device the resistivity of the thin Au film is measured for different values of an out of plane magnetic field. The magnetoresistance is then calculated as $(\rho_{B}-\rho_{B=0})/\rho_{B=0}$. The magnetoresistance plot for the sample with Au layer thickness of 8nm is shown in Figure \ref{fig:MR}. MR data show a clear signal of weak anti-localization. The effect of weak anti-localization grows weaker as the thickness of Au increases. This is expected since the higher the confinement of the electrons, the greater the chance of the electrons to follow a closed loop trajectory that will eventually contribute to either weak localization or weak anti-localization depending on the nature of wavefunction interference. The fact that weak anti-localization is observed can be attributed to the presence of strong spin orbit coupling in Au \cite{bergmann1982weak,bergman1982influence}. Interestingly enough, the magnitude of weak anti-localization has a very similar temperature scaling with $\Delta V$, at least for the thinner samples where a large and positive $\Delta V$ has been measured. The comparison between $\Delta V$ and $(\rho_{B_{max}}-\rho_{B=0})/\rho_{B=0}$ can be seen in Figure \ref{fig:DV_MR}.

\section{Discussion}
The reversal of the direction of spin polarization suggests that more than one spin polarization mechanism, with different spin polarization directions,  are present in the system. The low-thickness data points show an abrupt drop of $\Delta V$ with respect to the thickness of the Au film. This is consistent with the exponential decrease in spin polarization as a function of film thickness in the case of Rashba-Edelstein. On top of that, the sign of $\Delta V$ predicted from theory for the case of Rashba-Edelstein is positive which agrees with the sign of $\Delta V$ for samples of thickness 8~nm to 14~nm. The sign of $\Delta V$ is switched between 14~nm and 16~nm which indicates the inversion of the direction of spin polarization. For thicknesses 16~nm - 18~nm $\Delta V$ is negative and it is indistinguishable in magnitude from device to device. This similarity in the magnitude of the $\Delta V$ signal between the devices that display a negative $\Delta V$ is consistent with what is expected in the case when spin Hall effect is responsible for spin polarization which suggests that the amplitude of $\Delta V$ increases and saturates above a certain thickness of Au film.

Finally, we observe a similarity between the scaling of $\Delta V$ and $(\rho_{B}-\rho_{B=0})/\rho_{B=0}$ both with temperature and thickness (at least for 8~nm - 14~nm). This is likely due to the fact that both WAL and the two candidate mechanisms that can explain $\Delta V$ are a result of the strong SOC in Au.

\begin{acknowledgments}
The author is grateful to Dr. I. Appelbaum for supervising the project, providing the graphic in Fig. \ref{fig:E_f(k)}, and aiding in the theoretical derivations in Section II, and to Dr. P. Li for co-supervising the project. The author is thankful to the Maryland NanoCenter and its FabLab for providing their facilities.

This work was supported by the Office
of Naval Research under Contract N000141410317, and the
Defense Threat Reduction Agency under Contract HDTRA1-
13-1-0013.
\end{acknowledgments}

\newpage
\bibliography{main}

\begin{thebibliography}{15}%
\makeatletter
\providecommand \@ifxundefined [1]{%
 \@ifx{#1\undefined}
}%
\providecommand \@ifnum [1]{%
 \ifnum #1\expandafter \@firstoftwo
 \else \expandafter \@secondoftwo
 \fi
}%
\providecommand \@ifx [1]{%
 \ifx #1\expandafter \@firstoftwo
 \else \expandafter \@secondoftwo
 \fi
}%
\providecommand \natexlab [1]{#1}%
\providecommand \enquote  [1]{``#1''}%
\providecommand \bibnamefont  [1]{#1}%
\providecommand \bibfnamefont [1]{#1}%
\providecommand \citenamefont [1]{#1}%
\providecommand \href@noop [0]{\@secondoftwo}%
\providecommand \href [0]{\begingroup \@sanitize@url \@href}%
\providecommand \@href[1]{\@@startlink{#1}\@@href}%
\providecommand \@@href[1]{\endgroup#1\@@endlink}%
\providecommand \@sanitize@url [0]{\catcode `\\12\catcode `\$12\catcode
  `\&12\catcode `\#12\catcode `\^12\catcode `\_12\catcode `\%12\relax}%
\providecommand \@@startlink[1]{}%
\providecommand \@@endlink[0]{}%
\providecommand \url  [0]{\begingroup\@sanitize@url \@url }%
\providecommand \@url [1]{\endgroup\@href {#1}{\urlprefix }}%
\providecommand \urlprefix  [0]{URL }%
\providecommand \Eprint [0]{\href }%
\providecommand \doibase [0]{http://dx.doi.org/}%
\providecommand \selectlanguage [0]{\@gobble}%
\providecommand \bibinfo  [0]{\@secondoftwo}%
\providecommand \bibfield  [0]{\@secondoftwo}%
\providecommand \translation [1]{[#1]}%
\providecommand \BibitemOpen [0]{}%
\providecommand \bibitemStop [0]{}%
\providecommand \bibitemNoStop [0]{.\EOS\space}%
\providecommand \EOS [0]{\spacefactor3000\relax}%
\providecommand \BibitemShut  [1]{\csname bibitem#1\endcsname}%
\let\auto@bib@innerbib\@empty
%</preamble>
\bibitem [{\citenamefont {Li}\ \emph {et~al.}(2016)\citenamefont {Li},
  \citenamefont {van‘t Erve}, \citenamefont {Rajput}, \citenamefont {Li},\
  and\ \citenamefont {Jonker}}]{li2016direct}%
  \BibitemOpen
  \bibfield  {author} {\bibinfo {author} {\bibfnamefont {C.~H.}\ \bibnamefont
  {Li}}, \bibinfo {author} {\bibfnamefont {O.~M.}\ \bibnamefont {van‘t
  Erve}}, \bibinfo {author} {\bibfnamefont {S.}~\bibnamefont {Rajput}},
  \bibinfo {author} {\bibfnamefont {L.}~\bibnamefont {Li}}, \ and\ \bibinfo
  {author} {\bibfnamefont {B.~T.}\ \bibnamefont {Jonker}},\ }\href@noop {}
  {\bibfield  {journal} {\bibinfo  {journal} {Nat. Comm.}\ }\textbf {\bibinfo
  {volume} {7}},\ \bibinfo {pages} {13518} (\bibinfo {year}
  {2016})}\BibitemShut {NoStop}%
\bibitem [{\citenamefont {Ando}\ \emph {et~al.}(2014)\citenamefont {Ando},
  \citenamefont {Hamasaki}, \citenamefont {Kurokawa}, \citenamefont {Ichiba},
  \citenamefont {Yang}, \citenamefont {Novak}, \citenamefont {Sasaki},
  \citenamefont {Segawa}, \citenamefont {Ando},\ and\ \citenamefont
  {Shiraishi}}]{ando2014electrical}%
  \BibitemOpen
  \bibfield  {author} {\bibinfo {author} {\bibfnamefont {Y.}~\bibnamefont
  {Ando}}, \bibinfo {author} {\bibfnamefont {T.}~\bibnamefont {Hamasaki}},
  \bibinfo {author} {\bibfnamefont {T.}~\bibnamefont {Kurokawa}}, \bibinfo
  {author} {\bibfnamefont {K.}~\bibnamefont {Ichiba}}, \bibinfo {author}
  {\bibfnamefont {F.}~\bibnamefont {Yang}}, \bibinfo {author} {\bibfnamefont
  {M.}~\bibnamefont {Novak}}, \bibinfo {author} {\bibfnamefont
  {S.}~\bibnamefont {Sasaki}}, \bibinfo {author} {\bibfnamefont
  {K.}~\bibnamefont {Segawa}}, \bibinfo {author} {\bibfnamefont
  {Y.}~\bibnamefont {Ando}}, \ and\ \bibinfo {author} {\bibfnamefont
  {M.}~\bibnamefont {Shiraishi}},\ }\href@noop {} {\bibfield  {journal}
  {\bibinfo  {journal} {Nano Lett.}\ }\textbf {\bibinfo {volume} {14}},\
  \bibinfo {pages} {6226} (\bibinfo {year} {2014})}\BibitemShut {NoStop}%
\bibitem [{\citenamefont {Dankert}\ \emph {et~al.}(2015)\citenamefont
  {Dankert}, \citenamefont {Geurs}, \citenamefont {Kamalakar}, \citenamefont
  {Charpentier},\ and\ \citenamefont {Dash}}]{dankert2015room}%
  \BibitemOpen
  \bibfield  {author} {\bibinfo {author} {\bibfnamefont {A.}~\bibnamefont
  {Dankert}}, \bibinfo {author} {\bibfnamefont {J.}~\bibnamefont {Geurs}},
  \bibinfo {author} {\bibfnamefont {M.~V.}\ \bibnamefont {Kamalakar}}, \bibinfo
  {author} {\bibfnamefont {S.}~\bibnamefont {Charpentier}}, \ and\ \bibinfo
  {author} {\bibfnamefont {S.~P.}\ \bibnamefont {Dash}},\ }\href@noop {}
  {\bibfield  {journal} {\bibinfo  {journal} {Nano Lett.}\ }\textbf {\bibinfo
  {volume} {15}},\ \bibinfo {pages} {7976} (\bibinfo {year}
  {2015})}\BibitemShut {NoStop}%
\bibitem [{\citenamefont {Tang}\ \emph {et~al.}(2014)\citenamefont {Tang},
  \citenamefont {Chang}, \citenamefont {Kou}, \citenamefont {Murata},
  \citenamefont {Choi}, \citenamefont {Lang}, \citenamefont {Fan},
  \citenamefont {Jiang}, \citenamefont {Montazeri}, \citenamefont {Jiang} \emph
  {et~al.}}]{tang2014electrical}%
  \BibitemOpen
  \bibfield  {author} {\bibinfo {author} {\bibfnamefont {J.}~\bibnamefont
  {Tang}}, \bibinfo {author} {\bibfnamefont {L.-T.}\ \bibnamefont {Chang}},
  \bibinfo {author} {\bibfnamefont {X.}~\bibnamefont {Kou}}, \bibinfo {author}
  {\bibfnamefont {K.}~\bibnamefont {Murata}}, \bibinfo {author} {\bibfnamefont
  {E.~S.}\ \bibnamefont {Choi}}, \bibinfo {author} {\bibfnamefont
  {M.}~\bibnamefont {Lang}}, \bibinfo {author} {\bibfnamefont {Y.}~\bibnamefont
  {Fan}}, \bibinfo {author} {\bibfnamefont {Y.}~\bibnamefont {Jiang}}, \bibinfo
  {author} {\bibfnamefont {M.}~\bibnamefont {Montazeri}}, \bibinfo {author}
  {\bibfnamefont {W.}~\bibnamefont {Jiang}},  \emph {et~al.},\ }\href@noop {}
  {\bibfield  {journal} {\bibinfo  {journal} {Nano Lett.}\ }\textbf {\bibinfo
  {volume} {14}},\ \bibinfo {pages} {5423} (\bibinfo {year}
  {2014})}\BibitemShut {NoStop}%
\bibitem [{\citenamefont {Tian}\ \emph {et~al.}(2015)\citenamefont {Tian},
  \citenamefont {Miotkowski}, \citenamefont {Hong},\ and\ \citenamefont
  {Chen}}]{tian2015electrical}%
  \BibitemOpen
  \bibfield  {author} {\bibinfo {author} {\bibfnamefont {J.}~\bibnamefont
  {Tian}}, \bibinfo {author} {\bibfnamefont {I.}~\bibnamefont {Miotkowski}},
  \bibinfo {author} {\bibfnamefont {S.}~\bibnamefont {Hong}}, \ and\ \bibinfo
  {author} {\bibfnamefont {Y.~P.}\ \bibnamefont {Chen}},\ }\href@noop {}
  {\bibfield  {journal} {\bibinfo  {journal} {Sci. Rep.}\ }\textbf {\bibinfo
  {volume} {5}} (\bibinfo {year} {2015})}\BibitemShut {NoStop}%
\bibitem [{\citenamefont {Lee}\ \emph {et~al.}(2015)\citenamefont {Lee},
  \citenamefont {Richardella}, \citenamefont {Hickey}, \citenamefont
  {Mkhoyan},\ and\ \citenamefont {Samarth}}]{lee2015mapping}%
  \BibitemOpen
  \bibfield  {author} {\bibinfo {author} {\bibfnamefont {J.~S.}\ \bibnamefont
  {Lee}}, \bibinfo {author} {\bibfnamefont {A.}~\bibnamefont {Richardella}},
  \bibinfo {author} {\bibfnamefont {D.~R.}\ \bibnamefont {Hickey}}, \bibinfo
  {author} {\bibfnamefont {K.~A.}\ \bibnamefont {Mkhoyan}}, \ and\ \bibinfo
  {author} {\bibfnamefont {N.}~\bibnamefont {Samarth}},\ }\href@noop {}
  {\bibfield  {journal} {\bibinfo  {journal} {Phys. Rev. B}\ }\textbf {\bibinfo
  {volume} {92}},\ \bibinfo {pages} {155312} (\bibinfo {year}
  {2015})}\BibitemShut {NoStop}%
\bibitem [{\citenamefont {Hsieh}\ \emph {et~al.}(2009)\citenamefont {Hsieh},
  \citenamefont {Xia}, \citenamefont {Qian}, \citenamefont {Wray},
  \citenamefont {Dil}, \citenamefont {Meier}, \citenamefont {Osterwalder},
  \citenamefont {Patthey}, \citenamefont {Checkelsky}, \citenamefont {Ong}
  \emph {et~al.}}]{hsieh2009tunable}%
  \BibitemOpen
  \bibfield  {author} {\bibinfo {author} {\bibfnamefont {D.}~\bibnamefont
  {Hsieh}}, \bibinfo {author} {\bibfnamefont {Y.}~\bibnamefont {Xia}}, \bibinfo
  {author} {\bibfnamefont {D.}~\bibnamefont {Qian}}, \bibinfo {author}
  {\bibfnamefont {L.}~\bibnamefont {Wray}}, \bibinfo {author} {\bibfnamefont
  {J.}~\bibnamefont {Dil}}, \bibinfo {author} {\bibfnamefont {F.}~\bibnamefont
  {Meier}}, \bibinfo {author} {\bibfnamefont {J.}~\bibnamefont {Osterwalder}},
  \bibinfo {author} {\bibfnamefont {L.}~\bibnamefont {Patthey}}, \bibinfo
  {author} {\bibfnamefont {J.}~\bibnamefont {Checkelsky}}, \bibinfo {author}
  {\bibfnamefont {N.}~\bibnamefont {Ong}},  \emph {et~al.},\ }\href@noop {}
  {\bibfield  {journal} {\bibinfo  {journal} {Nature}\ }\textbf {\bibinfo
  {volume} {460}},\ \bibinfo {pages} {1101} (\bibinfo {year}
  {2009})}\BibitemShut {NoStop}%
\bibitem [{\citenamefont {Li}\ and\ \citenamefont
  {Appelbaum}(2016)}]{Li_PRB16}%
  \BibitemOpen
  \bibfield  {author} {\bibinfo {author} {\bibfnamefont {P.}~\bibnamefont
  {Li}}\ and\ \bibinfo {author} {\bibfnamefont {I.}~\bibnamefont {Appelbaum}},\
  }\href {\doibase 10.1103/PhysRevB.93.220404} {\bibfield  {journal} {\bibinfo
  {journal} {Phys. Rev. B}\ }\textbf {\bibinfo {volume} {93}},\ \bibinfo
  {pages} {220404} (\bibinfo {year} {2016})}\BibitemShut {NoStop}%
\bibitem [{\citenamefont {Bychkov}\ and\ \citenamefont
  {Rashba}(1984)}]{bychkov1984properties}%
  \BibitemOpen
  \bibfield  {author} {\bibinfo {author} {\bibfnamefont {Y.~A.}\ \bibnamefont
  {Bychkov}}\ and\ \bibinfo {author} {\bibfnamefont {E.}~\bibnamefont
  {Rashba}},\ }\href@noop {} {\bibfield  {journal} {\bibinfo  {journal} {JETP
  Lett.}\ }\textbf {\bibinfo {volume} {39}},\ \bibinfo {pages} {78} (\bibinfo
  {year} {1984})}\BibitemShut {NoStop}%
\bibitem [{\citenamefont {Rashba}(1959)}]{rashba1959symmetry}%
  \BibitemOpen
  \bibfield  {author} {\bibinfo {author} {\bibfnamefont {E.}~\bibnamefont
  {Rashba}},\ }\href@noop {} {\bibfield  {journal} {\bibinfo  {journal} {Soviet
  Physics-Solid State}\ }\textbf {\bibinfo {volume} {1}},\ \bibinfo {pages}
  {368} (\bibinfo {year} {1959})}\BibitemShut {NoStop}%
\bibitem [{\citenamefont {Rashba}\ and\ \citenamefont
  {Sheka}(1959)}]{rashba2015symmetry}%
  \BibitemOpen
  \bibfield  {author} {\bibinfo {author} {\bibfnamefont {E.}~\bibnamefont
  {Rashba}}\ and\ \bibinfo {author} {\bibfnamefont {V.}~\bibnamefont {Sheka}},\
  }\href@noop {} {\bibfield  {journal} {\bibinfo  {journal} {Fiz. Tverd. Tela:
  Collected Papers}\ }\textbf {\bibinfo {volume} {2}},\ \bibinfo {pages} {62}
  (\bibinfo {year} {1959})}\BibitemShut {NoStop}%
\bibitem [{\citenamefont {Yu}\ and\ \citenamefont
  {Flatt{\'e}}(2002)}]{yu2002electric}%
  \BibitemOpen
  \bibfield  {author} {\bibinfo {author} {\bibfnamefont {Z.}~\bibnamefont
  {Yu}}\ and\ \bibinfo {author} {\bibfnamefont {M.}~\bibnamefont
  {Flatt{\'e}}},\ }\href@noop {} {\bibfield  {journal} {\bibinfo  {journal}
  {Phys. Rev. B}\ }\textbf {\bibinfo {volume} {66}},\ \bibinfo {pages} {201202}
  (\bibinfo {year} {2002})}\BibitemShut {NoStop}%
\bibitem [{Note1()}]{Note1}%
  \BibitemOpen
  \bibinfo {note} {The charge current density was kept constant for all
  measurements at $j=2\times 10^9$ A/m$^2$. For the sample geometry in this
  experiment, this corresponds to a current of $200 \mu $A per nm of thickness
  in the Au thin film.}\BibitemShut {Stop}%
\bibitem [{\citenamefont {Bergmann}(1982)}]{bergmann1982weak}%
  \BibitemOpen
  \bibfield  {author} {\bibinfo {author} {\bibfnamefont {G.}~\bibnamefont
  {Bergmann}},\ }\href@noop {} {\bibfield  {journal} {\bibinfo  {journal}
  {Solid State Comm.}\ }\textbf {\bibinfo {volume} {42}},\ \bibinfo {pages}
  {815} (\bibinfo {year} {1982})}\BibitemShut {NoStop}%
\bibitem [{\citenamefont {Bergman}(1982)}]{bergman1982influence}%
  \BibitemOpen
  \bibfield  {author} {\bibinfo {author} {\bibfnamefont {G.}~\bibnamefont
  {Bergman}},\ }\href@noop {} {\bibfield  {journal} {\bibinfo  {journal} {Phys.
  Rev. Lett.}\ }\textbf {\bibinfo {volume} {48}},\ \bibinfo {pages} {1046}
  (\bibinfo {year} {1982})}\BibitemShut {NoStop}%
\end{thebibliography}%
\end{document}